# 3D surface profiling via photonic integrated geometric sensor

Ziyao Zhang[1,†], Yizhi Wang[1,†], Chunhui Yao[1,3,†], Huiyu Huang[1], Rui Ma[1], Xin Du[2], Wanlu Zhang[1], Zhitian Shi[1], Minjia Chen[1], Ting Yan[3], Liang Ming[3], Yuxiao Ye[3], Richard Penty[1], Qixiang Cheng[1,3,*]

[1] Electrical Engineering Division, Department of Engineering, University of Cambridge, UK
[2] The cavendish laboratory, Department of Physics, University of Cambridge, Cambridge, UK
[3] GlitterinTech Limited, Xuzhou, China

[†]*These authors contributed equally.*
[*]*Corresponding author: qc223@cam.ac.uk*

## Abstract

Measurements of microscale surface patterns are essential for process and quality control in industries across semiconductors, micro-machining, and biomedicines. However, the development of miniaturized and intelligent profiling systems remains a longstanding challenge, primarily due to the complexity and bulkiness of existing benchtop systems required to scan large-area samples. A real-time, in-situ, and fast detection alternative is therefore highly desirable for predicting surface topography on the fly. In this paper, we present an ultracompact geometric profiler based on photonic integrated circuits, which directly encodes the optical reflectance of the sample and decodes it with a neural network. This platform is free of complex interferometric configurations and avoids time-consuming nonlinear fitting algorithms. We show that a silicon programmable circuit can generate pseudo-random kernels to project input data into higher dimensions, enabling efficient feature extraction via a lightweight one-dimensional convolutional neural network. Our device is capable of high-fidelity, fast-scanning-rate thickness identification for both smoothly varying samples and intricate 3D printed emblem structures, paving the way for a new class of compact geometric sensors.

## Introduction

Accurate, high-resolution measurement of 3D surface topography is essential for a wide range of applications, particularly those involving complex and finely detailed structures such as precision manufacturing, microelectronics, biomedical engineering, and soft robotics (*1*, *2*). Spectral sensing-based optical metrology offers an in-situ, non-contact solution for surface characterization and is widely adopted across industrial and scientific domains (*3*). However, conventional benchtop approaches—such as white-light interferometry and fringe projection profilometry—typically depend on interferometric setups with high-performance spectrometers to extract surface information from fringe patterns (*4*, *5*), as shown in Fig. 1(a). These systems often involve bulky free-space optics, demand precise alignment, and operate under tight mechanical tolerances. Furthermore, they follow a multi-step processing pipeline—comprising of detection, signal recovery, preprocessing, and prediction—which introduces considerable computational overhead. Such constraints limit their suitability for fast, on-the-fly measurements in mobile or handheld sensing scenarios (*6*). Additionally, the need to record user-specific spectral information may raise privacy concerns.

Recent years have witnessed the surge in deep learning research and applications (*7*), opening up new possibilities for interpreting unconventional and unstructured datatypes often beyond the capabilities of traditional machine learning algorithms. At the core of neural networks and models such as reservoir computing lies the principle of complex connectivity, which enables the mapping of input data into high-dimensional function spaces (*8*). In these transformed spaces, features that are not directly discernible in the original input can become separable. Moreover, with appropriate training, these models can develop inherent robustness to natural variations—making them well-suited for handling fabrication tolerances and noise commonly encountered in real-world sensing scenarios. Among these models, deep neural networks—particularly convolutional architectures—stand out as ideal decoders for end-to-end geometric sensing, offering a direct mapping from raw optical signals to physically meaningful quantities (*9*).

In parallel, rapid advances in photonic integration technologies have enabled the development of miniaturized, cost-effective optical sensors for a wide range of metrological applications (*10*, *11*). Notable examples include the directional Huygens dipole-based displacement sensor (*12*), the Bessel–Gaussian beam generator for rotational Doppler sensing (*13*), and integrated spectrometers for material and biomarker analysis (*14, 15*). Among these, computational photonic systems—featuring specially engineered sampling circuits that produce feature-rich responses—have demonstrated exceptional performance in capturing complex, information-dense signals (*16*). This emerging design paradigm, which co-optimizes physical encoding and algorithmic decoding, provides a compelling foundation for the development of high-performance integrated geometric sensors.

Building on this concept of efficient encoder-decoder architecture, we propose a novel class of on-chip geometric profiler for detecting surface micro-patterns. As illustrated in Fig. 1(b), our system eliminates the need for complex interferometric setups, relying instead on a photonic integrated chip equipped with a single-channel programmable sampling circuit. This circuit performs pseudo-random spectral sampling of the light reflected from the sample surface. The resulting encoded spectral signals are then decoded using a one-dimensional convolutional neural network (1D-CNN), which reconstructs the surface topography with high fidelity. Experimentally, our device achieved high-fidelity, fast-scanning-rate thickness identification for both smoothly varying samples and intricate 3D printed emblem structures. Owing to its highly streamlined detection-to-prediction workflow, the proposed sensor bypasses intermediate reconstruction and data storage, reducing computational overhead to a single inference pass. With the integration of faster modulation techniques and improved mechanical components, the system's throughput can be significantly enhanced further. In summary, our design provides a compact, alignment-free, and data-driven alternative to conventional optical metrology systems, enabling accurate, lightweight, and real-time 3D geometric reconstruction.

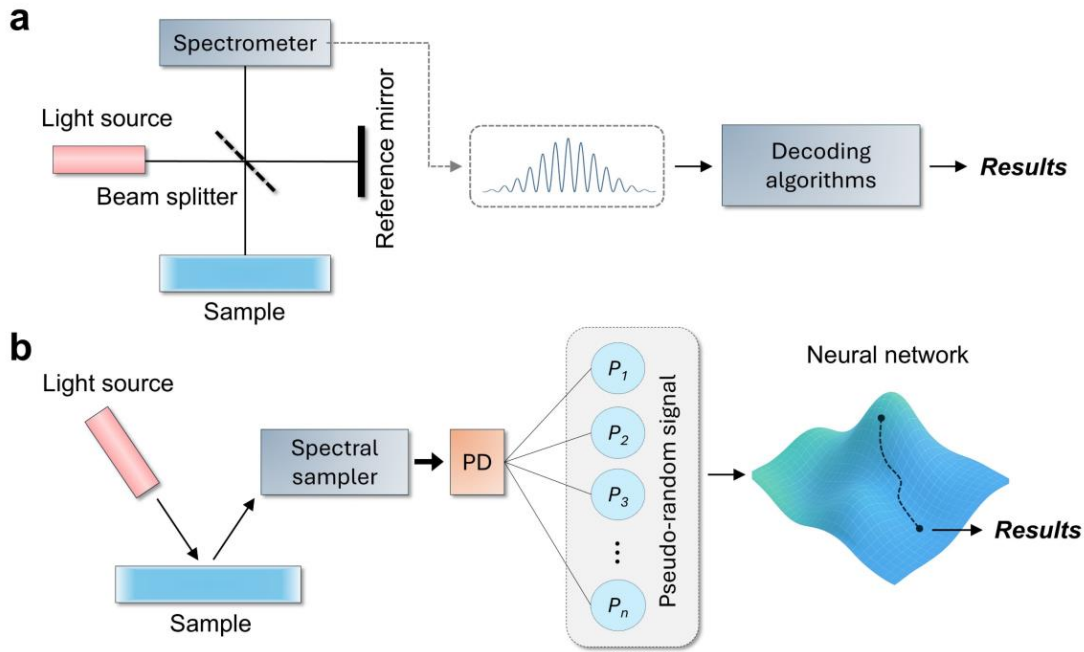

**Fig. 1.** Schematic illustrations of (a) conventional benchtop interferometric spectrometry systems and (b) our integrated on-chip geometric sensing scheme, respectively.

## Results

### Design of Photonic Encoder for Geometric Sensing

The basis of geometric sensing can be described as a mapping between the sampled features' $m$-dimensional function space and the output features' $n$-dimensional function space. In many practical applications, the extractable physical parameter is of modest size (*17*). For example, in surface profiling, the output dimension $n$ is effectively one, corresponding to the sample's thickness/height. This sets up a scenario where $m \gg n$ for the encoder-decoder pair, making it plausible for the desired feature to be directly extracted from the sampled information. Nonetheless, there are two key requirements for the geometric sensing system: 1) the design of a complex enough encoder to collect as much information as possible from the scanned sample, and 2) a decoder capable of predicting thickness directly. Following the essence of compressed sensing, it is preferable for each sampled channel to contain more information than a plain one-to-one direction. In other words, for retrieval purposes, the channel reading of intensity mixing between different wavelengths contains more information than a single PD reading at a single wavelength. Thus, a randomized kernel is commonly employed for collecting sample's rich spectrum information (*18*, *19*).

To physically realize a compact and information-rich encoder, we implement a programmable photonic integrated circuit capable of generating a large number of sampling channels ($m$), as shown by Fig. 2(a). To enable rich and dense spectral content, our circuit comprises six over-coupled microring resonators (MRRs), each designed with a varying perimeter to produce a unique free spectral range (FSR). Thus, the cascade of these MRRs yields a wavelength-selective, pseudo-random spectral response. Meanwhile, each MRR is integrated with a thermo-optic (T-O) phase shifter, enabling the circuit to sequentially generate distinct sampling states by dynamically tuning the phase configuration. These temporally modulated responses act as spectral encoders to capture the rich geometric information embedded in the reflected light from the sample surface (see Fig. 1(b)). More design details are provided in Supplementary Section 1.

Figure 2(b) further illustrates the overall encoding workflow, with the insets showing some representative sampling responses and corresponding intensity readings measured from the photodetector (PD). The designed sampling circuit generates ultra-dense spectral fluctuations with sharp roll-offs. In practice, a total set of 729 distinct sampling responses were produced by assigning three discrete phase states to each of the six MRRs (i.e., $3^6$ combinations), providing a balanced trade-off between encoding performance, measurement time, and computational cost (see relevant discussions in Supplementary Section 2).

**Surface Profiling Decoder Based on 1D-CNN**

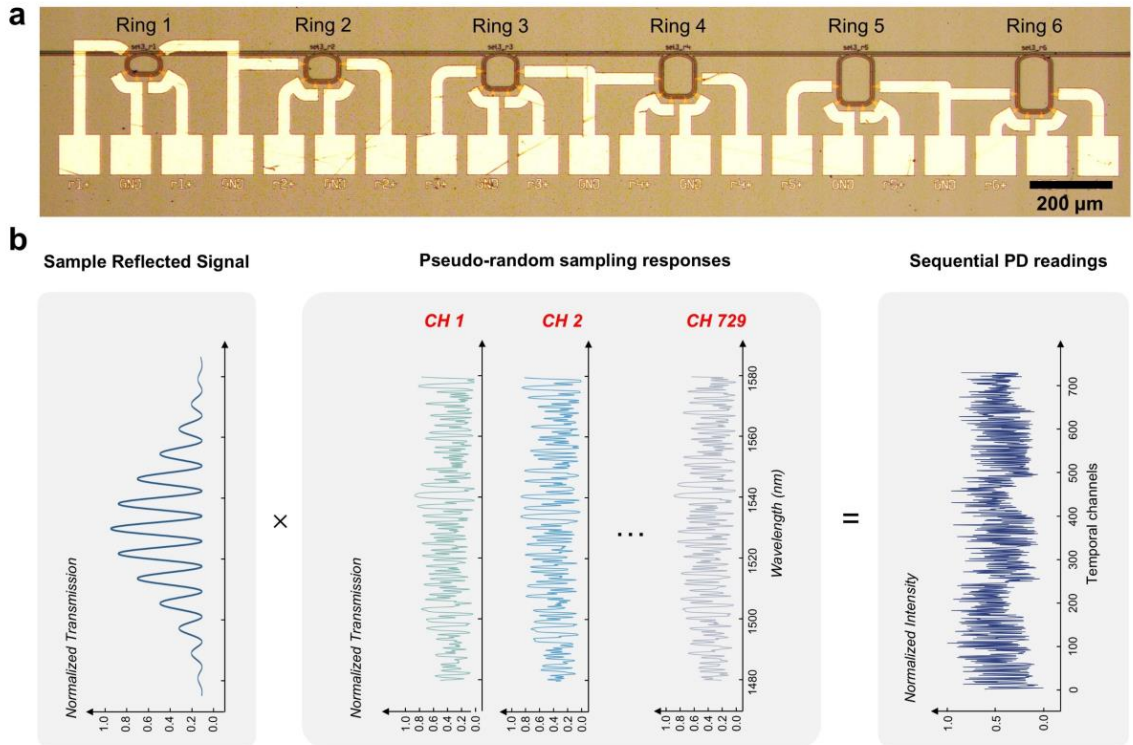

**Fig. 2.** (a) Microscopic photograph of the integrated spectral sampler. (b) Workflow diagram of the PIC-based sampling process for geometric sensing, including representative temporally modulated pseudo-random spectral responses and the output intensity readings measured at the photodetector.

With the PIC-based sampler as the encoder, the reflected sample signal is passed through a pseudo-random transmission matrix for each configuration. Each configuration gives rise to a single PD value in the data input vector, pending computation with the decoder. To complete the encoder-decoder pair, we employ a model based on one-dimensional convolutional neural network (1D-CNN). As indicated by Fig. 3(a), the model consists of two parts consisting of the convolutional layers and the fully connected forward layers. (See Methods for details on the model construction.)

Aligning with the essence of the geometric profiler to be a physically informed construction, the choice of the model architecture is also based on the physical interpretation of the task. Since the network input is the results of discretely defined mask based on the voltage setting of 729 variations, the reconfiguration command is acted upon each ring sequentially. (See Fig.2 (d) for combined transmission matrix of the PIC sampler). While the unbalanced ring structure in each stage aims to provide a pseudo-random mask, the effect of the stepwise modulation on the ring should be more traceable than multiple varying controls.

Equivalently speaking, two transmission masks with setting difference in one ring are more likely to exhibit some level of relevance than two masks with setting differences in multiple rings. Therefore, based on our channel definition, it is expected that channels grouped closely together are more likely to share physical feature resemblance than channels which are very far apart. Unlike the typical reconstruction process, where the extracted information can be solely derived with the combination information, spatial relevance in channels is physically relevant for our geometric sensor.

In addition, while the channel readings might exhibit spatial importance, they shouldn't bear any direct physical temporal importance. To expand on this, defining the first channel to be no voltage on all rings is physically equivalent to starting with all rings at maximum voltages, provided that the voltage changing logic is the same for both cases.

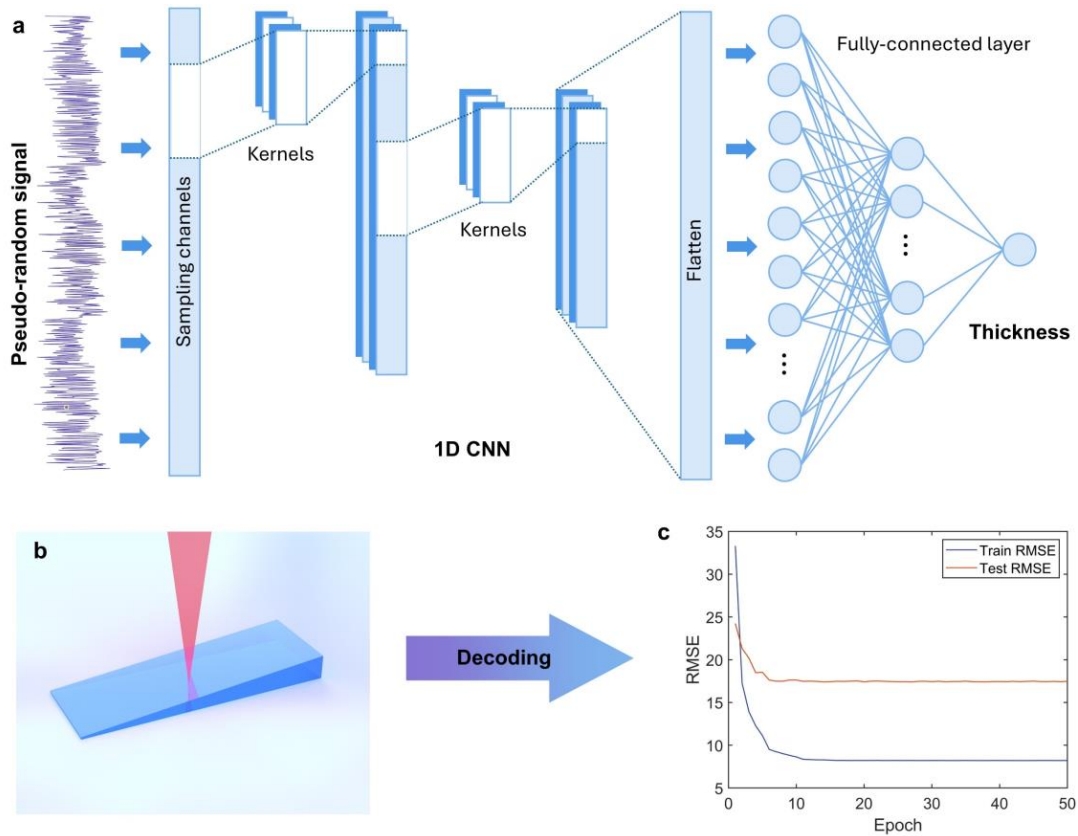

**Fig. 3.** (a) Pseudo-random spectral signals are fed into a 1D-CNN model for direct regression of sample thickness. (b) Surface topology of the slope sample to build the training database. (c) Training and testing RMSE of the 1D-CNN network versus training epochs.

Combining these physical observations, we choose a one-dimensional convolutional neural network as the core construction of our decoder network. In our construction, we hypothesize that the CNN part of our model can project the sample PD readings into higher dimensional spaces. This means that a further prediction/numerical estimation part is required for bringing the projected data back to an interpretable thickness prediction. Thus, following the CNN part, we employ a MLP structure with reducing neuron sizes in each layer until the single neuron at the output layer for matching the target of a single thickness value. Further discussion on the model construction can be found in Methods and Supplementary Section 3.

**Data Acquisition and Training Strategy**

The construction of an appropriate database is key to success in training. With the goal of the thickness profiler in mind, we 3D printed a smooth varying slope-like sample as the data source. The thickness is varied between 20 to 200 µm. To recreate the proposed scenario of thickness prediction without reference arm, multiple samples with varying distance from the slope are taken for each slope height. A combined dataset of more than 90,000 data points is collected as the training dataset. An additional 2000 data points are reserved as the testing dataset. With each input sample, we used standard benchtop spectrometer to retrieve the target thickness of the structure. In each data pair, thickness is the sole parameter to optimize for based on a simple MSE loss function (see Fig.3 (a,b)). The details of the experimental set up for this data acquisition are provided in Supplementary Section 4.

With our specifically built database, we proceed to train the CNN decoder, as shown in Fig.3. Through training, both the training and testing datasets show converging behavior. The train RMSE converges to under 8.2 $\mu m$ at the end of 50 epochs, and roughly 17.5 $\mu m$ for test RMSE. As expected due to natural variations in unseen data, the testing RMSE is slightly higher than that of the training RMSE. The comparable error levels indicate effective training without overfitting. Considering the simplicity in system construction and the eliminated reconstruction process, these performances build a solid foundation for the geometric sensor as fast samplers.

**Experimental Validation and Analysis**

Figure 4 illustrates the geometric sensing system constructed using our PIC sampler. A superluminescent diode (SLD) is employed as the broadband light source. The emitted light, after polarization control, is routed through an optical circulator and launched into free space via a collimator. An objective lens focuses the beam onto the surface of the target sample. The reflected optical signal is collected by the same objective lens, recoupled into the fiber, and directed to the PIC sampler for spectral encoding. A high-speed driving board with a microcontroller unit (MCU) is incorporated to enable programmable modulation of the PIC.

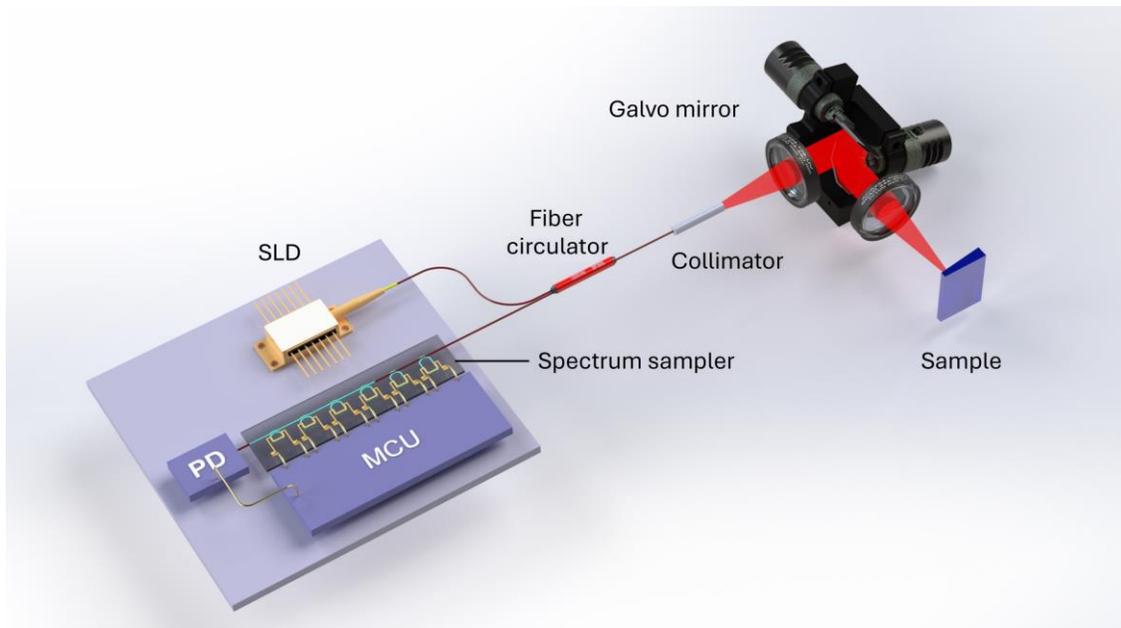

**Fig. 4.** Schematic of the experimental setup for geometric sensing using our integrated PIC sampler.

To validate the robustness of our sensor, we employ two demonstration routes: 1) a repeatability test using a smoothly varying slope sample, and 2) identification of unseen, sharply varying discrete emblem patterns. To ensure broad generalizability of our profiler, we fabricated an additional slope-shaped sample in a separate printing batch. The pre-trained network model was used to predict the slope height across the entire surface area of 1900 μm × 396 μm — extending beyond the one-axis movement used during training data acquisition. Sensor focusses displacement was controlled via a galvo mirror, with fine adjustments made relative to the focal plane. Figure 5 (a) shows the predicted topography of the test sample. Statistical analysis comparing the predicted and expected slope profiles indicates that most residual deviation fall within ±30 μm, following a Gaussian-like distribution centered around zero (see Fig.5 (b)), with an average deviation of 18.5 μm. These performance metrics align well with those observed during training. The consistent identification of separately fabricated structures demonstrates the robustness of the geometric profiler, with no observable systematic variation across different fabrication batches. One limiting factor in our current experimental setup is the reliance on moving mechanical parts and free-space optical components. Further improvements in sensor performance may be achieved by upgrading the complementary system associated with the PIC encoder.

Sharp step-change profiles are commonly encountered in industrial applications such as thin-film coating. While the previous validation focused on the sensor's ability to reconstruct smoothly varying surface topographies, we now evaluate its performance in resolving discrete height levels with complex spatial patterns. For this purpose, a further evaluation of the sensor's ability to resolve discrete height levels with complicated spatial patterns, we selected a lion-shaped emblem as the test sample (see Fig.5 (c) for the CAD model structure). The emblem was fabricated using 3D printing with two defined thickness levels of 40 μm and 150 μm. The entire pattern spans a lateral area of 1000 μm × 1000 μm (see microscope photos in Fig.5 (d)). The steep sidewalls between regions highlight the presence of sharp thickness transitions. Scanning was performed with a resolution of 100 × 100 pixels, yielding a total of 10,000 individual deep-sensing inferences.

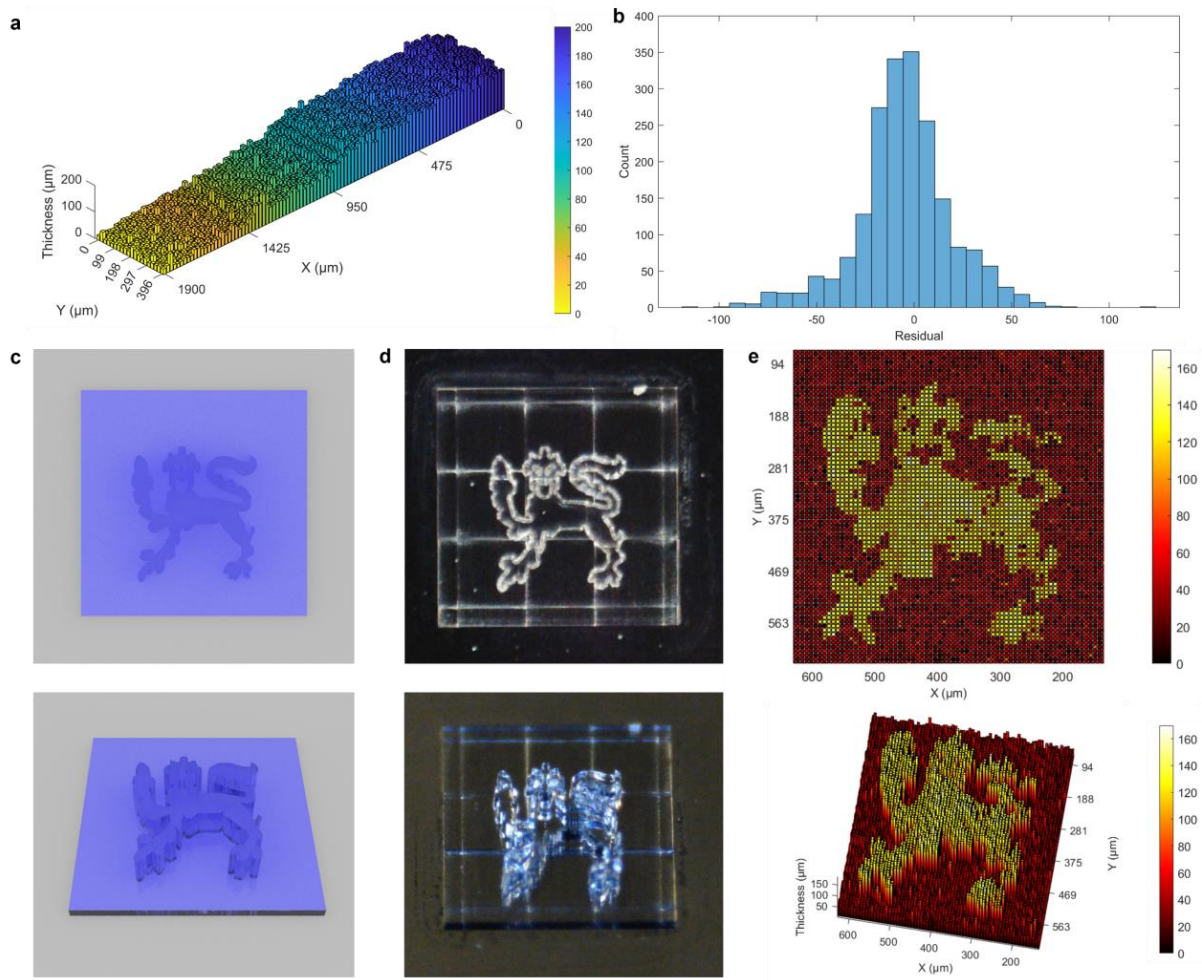

**Fig. 5**. (a) The predicted topography of a separate slope sample. (b) Histogram indicating the variations from the target height value. (c) The CAD model of proposed 'lion' emblem pattern. (d) Photo of the 3D printed emblem with size of 1000 μm × 1000 μm. (e) Geometric sensor prediction of the sample topography.

The predicated 3D profile is shown in Fig.5 (e), which includes both the 2D thickness map and the reconstructed 3D surface topography. The two thickness levels are clearly resolved, and the spatial boundaries of the lion pattern are accurately delineated, demonstrating high classification fidelity and strong topographic contrast. Distinct features of the emblem—such as the eyes and mouth of the lion—are identifiable in the predicted profile. For instance, the right eye spans approximately three pixels (30 μm), indicating good lateral resolution for capturing sharp, discrete surface transitions. In this experiment, the achievable lateral resolution primarily constrained by the mechanical precision of the galvo mirror and the focal spot size of the illumination source. Further improvements in resolution and sensing quality could be achieved through the integration of higher-precision components and optimized optical focusing systems.

## Conclusion

We present a novel geometric sensor featuring a simplified system architecture and decoder network. The spectrum encoder, based on a MRR array integrated on a photonic chip, enables straightforward control of encoded interference signals across 729 spectral channels. A 1D-CNN model is employed to decode the projected data and directly predict sample thickness. In this paper, we demonstrated the sensor's capability for topographic prediction on both smoothly varying and discretely structured samples. Testing on a sloped sample yielded an average thickness deviation of only 18.5 μm. Furthermore, a scan of a 3D lion emblem highlights the sensor's capability in achieving high lateral resolution, resolving features as small as 30 μm across.

The entire encoding process is implemented on a compact photonic chip with a footprint of just 0.4 mm², requiring no free-space alignment and significantly reducing system complexity. The deep-learning-based decoder traditional, cumbersome optical reconstruction algorithms, enabling real-time data processing. Given the high accuracy achieved with the current configuration, further enhancements in optical coupling and signal-to-noise ratio—such as the use of upgraded free-space optics—are expected to further minimize residual errors.

Looking ahead, the lightweight geometric sensor presents a promising pathway toward fully portable, real-time surface profilers. Future efforts will focus on integrating on-chip light sources and detectors to realize standalone profilometry modules, as well as extending the platform to support additional sensing modalities, such as multi-wavelength or spectrally multiplexed measurements. In summary, this work establishes the feasibility of combining integrated photonic encoding with neural-network-based decoding for high-speed, high-precision surface profiling, laying the groundwork for intelligent, miniaturized sensing systems.

## Materials and Methods

### Chip fabrication, packaging and control

The photonic chip is fabricated on a silicon nitride (SiN) platform through a CORNERSTONE MPW run and features a compact footprint of 0.4 mm². It is wire-bonded for electrical fan-out and optically interfaced via edge coupling with a lensed polarization-maintaining fiber (PMF). The chip is co-packaged with a thermistor and a thermoelectric cooler (TEC) for active temperature control, thereby maintaining thermal stability during operation. To enable phase modulation across individual MRRs, an automatic electrical driving board was developed, where a microcontroller unit (MCU) is programmed to generate control signals based on a pre-calibrated voltage look-up table. These signals are sent to a multi-channel digital-to-analog converter (DAC), whose analog outputs are subsequently amplified by dedicated circuits and applied to the photonic chip to achieve dynamic temporal sweeping of the sampling channels.

### Decoder network

In this work, the decoder network construction consists of 1D-CNN kernels and fully connected forward MLPs. The formulation of the 1D convolution can be described by Eq. (M1) for the input-output pair of $A$ and $B$ at layer $[l]$.

$$B^{[l]} = C^{[l]} + \sum K^{[l]}(i) \otimes A^{[l]}(i) \tag{M1}$$

The output is a direct Hadamard product between the weight matrix and the input levels, where $i$ denotes the starting index of the one-dimensional input vector. For the inference process, the kernel is static, but subject to optimization during the training process. The MLP layers are essentially direct all-to-all connectivity between neurons of successive layers. The mathematical expression for one forward layer is given in Eq. (M2).

$$A^{[l+1]} = f(W^{[l]}A^{[l]} + C^{[l]}) \tag{M2}$$

$W^{[l]}$ denotes the connectivity/weight matrix between layers, with the addition of a bias vector $C^{[l]}$. The net output is passed through a nonlinear activation function before entering the next layer.

For the standard construction of CNNs, nonlinear activation and pooling layers follow the convolution process. We also employed a batch normalization layer directly following the convolution layer. For the detailed description of the model structure, we denote each convolution layer as c[input channel, output channel, kernel size, stride, padding]. And the fully connected layer as m[input channel, output channel].

Starting from the sample input of PD readings, the single reading for each channel is projected by a layer of c1[1, 128, 3, 1, 1]. The two following layers are c2[128, 256, 3, 1, 1] and c3[256, 512, 3, 1, 1]. Between the layers, a batch normalization is applied before passing through a ReLU activation function. Following the activation, we used a pooling layer with kernel size of 2 and stride of 2.

In connection with the characteristics of the channels, we choose a small kernel width of 3. As mentioned, two consecutive channels might exhibit physically interpretable relations, but channels far apart should ideally be independent.

The output from the c3 layer is flattened and connected to subsequent fully connected layer of m1[46592, 256], m2[256, 128], and m3[128, 1]. The network is trained with backpropagation using Adam optimizer using initial learning rate of $1 \times 10^{-4}$ and decay rate of $1 \times 10^{-5}$.

**Fabrication of 3D-nanoprinted samples**

The sample is fabricated via 3D nano-printing technology enabled by Nanoscribe Photonics Professional GT2. Taking advantage of two-photon polymerization (2PP), the sample can be printed with sub-micron accuracy on an ITO coated silica substrate in a few minutes. A combination of 25x objective and IP-n162 resin is used here fors the balanced printing speed and quality. Slicing and hatching distances are set to be 0.2μm and 0.1μm respectively. After the printing process, the sample goes through a development process of 15 mins immersion in PGMEA and 5 mins immersion in IPA. UV curing is also applied to the sample to enhance the mechanical robustness after the development process is done.

## Acknowledgement

**Funding:** This work was supported by the European Union's Horizon 2020 research and innovation programme, project INSPIRE (101017088), and UK EPSRC, project QUDOS (EP/T028475/1).

## Author contributions

Z.Z. conceived the sensing system design, performed the data acquisition and experimental validation. Y.W. analysed the data with X.D.'s contribution. C.Y. designed the spectral sampler and drawn the chip layout, with T.Y., L.M., Y.Y.'s assistance. H.H fabricated the samples with Z.S.'s help. W.Z., C. Y. and M.C. built the chip testing set up. Z.Z. drafted the manuscript together with C.Y., Y.W., R.M. and Q.C. R.P. and Q.C. supervised the project.

## Competing interests

The authors declare no competing interests.

## Data and material availability

All data needed to evaluate the conclusions of this study are included in the main text and/or the Supplementary Materials. Additional data are available from the authors upon reasonable request.